\documentclass[twocolumn]{article}

\usepackage{amsmath}
\usepackage{gensymb}
\usepackage{subcaption}

\usepackage{authblk}

\usepackage{ulem} % allow the strikethrough command \sout
\usepackage{color}

\usepackage{graphicx}
\title{Synergistic Sensing: Application of SiNWs-PANI:MO$_x$ Heterostructures for Human Respiratory Monitoring}

\author[1]{M. T. Sultan}
\author[2]{A. Dumitru}
\author[1]{E. A. Fakhri}
\author[1]{R. E. Brophy}
\author[3]{S. T. Ingvarsson}
\author[1]{A. Manolescu}
\author[1]{H. G. Svavarsson}

\affil[1]{Reykjavik University, Dept. of Engineering, Menntavegur 1, 102 Reykjavik, Iceland}
\affil[2]{University of Bucharest, Faculty of Physics, PO Box MG-11, 077125, Magurele, Romania}
\affil[3]{Science Institute, University of Iceland, Dunhaga 3, 107 Reykjavik, Iceland}
\date{}

\begin{document}
%\revised{\today}%

\twocolumn[
 \begin{@twocolumnfalse}
 
\maketitle
 
\begin{abstract}
\noindent 

In this study we investigate novel hybrid structure of silicon nanowires (SiNWs) coated with PANI:metaloxide(MO$_x$) nanoparticles i.e., WO$_3$ and TiO$_2$. The SiNWs were fabricated using MACE, whereas PANI :MO$_x$ were deposited using chemical oxidative polymerization method on SiNWs. To this date little attempts has been done to utilize such hybrid structures for respiratory sensing. The structures were characterized using RAMAN spectroscopy, X-ray diffraction, Electron disperssive spectroscopy, and Scanning electron microscopy. The electrical characterization to obtain respiratory sensing reveals excellent response compared to those obtained for SiNWs:MO$_x$ and SiNWs:PANI. Such enhancement in sensitivity is attributed to formation p-n heterojunction along side with wider conduction channel provided of PANI, increased porosity in SiNWs/PANI:WO$_3$ hybrid structures, providing active sites, increased oxygen vacancies and large surface area compared to that of pure MO$_x$ nanoparticles. Further, an improved drift in base line and sensor stability was established  for the structure with PANI:WO$_3$ as compared to the PANI:TiO$_2$.\\\\

\end{abstract}

\begin{small}
\end{small}

\vspace{8mm}

\end{@twocolumnfalse}
]

%\pacs{}
%\keyword{???,????,????}

%\maketitle

%%%%%%%%%%%%%%%%%%%%
\section{Introduction}
%%%%%%%%%%%%%%%%%%%%
Breathing is a vital physiological process in living organisms. For humans, this process involves inhaling air containing oxygen into the lungs, where gas exchange occurs across the alveolar-capillary membrane. Carbon dioxide is excreted during exhalation, released through the nose or mouth. The entire process, from inhalation to exhalation, is known as the breathing or respiration cycle. Respiratory rate, a key vital sign, is used to monitor the progression of illness, with abnormal rates serving as critical markers of serious conditions.\par

Continuous monitoring of respiratory rate is crucial in various medical settings. For instance, substantial evidence indicates that alterations in respiratory rate can predict potentially serious clinical events, such as cardiac arrest or admission to intensive care units. Studies have shown that respiratory rate is a more reliable indicator than other vital measurements, such as pulse and blood pressure, in differentiating between stable patients and those at risk  \cite{Akbari,Yu2023}. Changes in respiratory rate measurements can identify high-risk patients up to 24 hours before an event, with a specificity of 95\%. Additionally, in the context of COVID-19, respiratory monitoring is critical for evaluating pulmonary function.\par

\begin{figure*}[htb]
\centering
\includegraphics[width=1\textwidth]{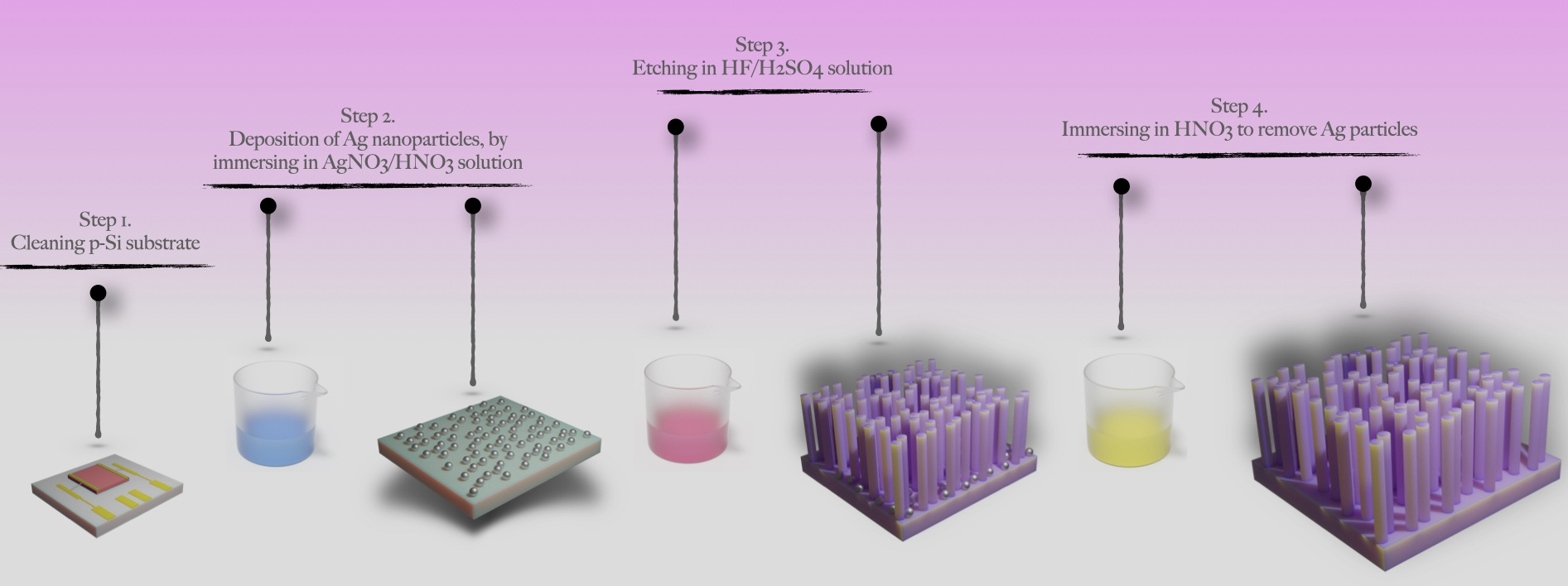}
\caption{Schematic representation of randomly aligned vertical SiNWs using MACE.}
\label{1}
\end{figure*}

Another important study involves, sleep apnea hypopnea syndrome (SAHS), which is a prevalent and potentially serious disorder characterized by at least 30 apneas during a normal 7-hour sleep, where each apnea may last at least for 10 seconds. SAHS leads to sleep interruptions, resulting in various pathophysiological changes, including complications such as hypertension and stroke. SAHS has become the second leading cause of stroke. Therefore, respiratory monitoring of individuals with SAHS is particularly important and urgent \cite{Andreozzi2021}.\par

One effective strategy for respiratory monitoring involves using the moisture content in exhaled human breath, where the relative humidity is about 60-88\% \cite{Mansour2020}. In this context, devices with active materials at the nano-scale have emerged as promising candidates for gas sensing applications due to their high surface-to-volume ratio and small physical dimensions, comparable to the charge screening length. A good gas sensor must exhibit high sensitivity and selectivity towards specific gases. Additionally, the sensor should offer long-term stability, repeatability, a low operating temperature, and consequently low power consumption. Furthermore, a cost-effective fabrication process is essential from an industrial perspective.\par
Nanostructures, for instance nanotubes \cite{Guo2022}, nanoparticles \cite{Kong2022, Sun2012}, nanosheets \cite{Ji2016, Umar2017}, and nanowires \cite{Comini2020, Akbari, Fakhri2023}, have shown good sensitivity to different gases. Among these nanostructures, silicon nanowires (SiNWs) have demonstrated substantial advantages \cite{Akbari}. SiNWs can be processed using relatively standard techniques, which allows for integration with standard complementary metal-oxide-semiconductor (CMOS) processes for very large-scale production. Additionally, SiNWs offer flexible doping concentrations and can be chemically functionalized for the selective detection of molecules in the gas phase. This makes SiNWs particularly suitable for advanced gas sensing applications.\par

Incorporating semiconductor metal oxides (MO$_x$) provides high electron mobility, high sensitivity, fast response/recovery time, long-term stability, and stable chemical properties  \cite{He2020, Masuda2023, Kulkarni2019}. Additionally, these materials allow easy adjustment of surface properties. In recent years, n-type semiconducting metal oxides, such as iron oxide, zinc oxide, titanium oxide, tin oxide, tungsten oxide and indium oxide  \cite{Bandgar2015, Bonyani2023, Pippara2021, Tian2016, Kulkarni2019, Masuda2023, Kong2022}, have become pivotal in gas sensing due to their low cost, high sensitivity, simplicity, and ease of integration into electronics \cite{He2020, Kulkarni2019}. However, gas sensors based on metal oxides (MO$_x$) face drawbacks like high resistance, high operating temperatures, and low selectivity \cite{He2020, Masuda2023, Kulkarni2019, Kumar2016}.\par

To address these issues, conducting polymers such as polyacetylene, polypyrrole, polyaniline (PANI), and poly-diacetylene have been explored \cite{Jain2023, Khuspe2012, Huang2011, Fratoddi2015}. These materials offer high conductivity, low-temperature operation, and low power consumption. Among them, PANI has been extensively studied and widely applied \cite{He2020} due to its efficiency in gas sensing, low cost, easy synthesis, room temperature operation, and low power consumption. Nevertheless, conducting polymer-based gas sensors suffer from poor stability, selectivity, and long response times \cite{Khuspe2012, Huyen2011, Kumar2016, Tian2016}.\par

Hybrid nanocomposites of metal oxides and conducting polymers present a promising solution to enhance the gas sensing properties of the individual components while maintaining their unique desirable properties. Recent advancements have seen the successful development of PANI-based hybrid nanocomposites by several research groups for detecting a variety of target gases \cite{Bonyani2023, He2020, Huyen2011, Kulkarni2019, Bandgar2015, Li2018}.\par

Therefore, we propose a novel SiNW-based  PANI:metal-oxide hybrid nanocomposite structure for respiratory sensing. In this work, SiNWs were synthesized using metal-assisted chemical etching, followed by coating with an organic-inorganic sensing layer comprising PANI and MO$_x$ nanoparticles (MO$_x$-NPs). This approach aims to leverage the combined advantages of both materials, potentially leading to significant improvements in respiratory sensing capabilities.\\

%%%%%%%%%%%%%%%%%%%%
\section{Experimental}
%%%%%%%%%%%%%%%%%%%%
%\subsection{Methodological Framework}
\subsection{Synthesis of SiNWs}
Synthesis of arrays of random SiNWs were carried out by applying metal (silver, Ag) assisted chemical etching (MACE) on \textit{p}-type $10\times 10$ mm$^2$ single-side polished Si-substrate of 525~µm thick, with the resistivity $\rho$ of 0.1-0.5~$\Omega$cm and 0.009 $\Omega$cm. The process steps of the synthesis (shown schematically in Fig. \ref{1} are as follows: 
\begin{itemize}
\item Deposition of Ag-NPs by immersing Si-substrates in a solution of 3 M HF and 1.5 mM AgNO$_3$ for 60 s, followed by rinsing in DI-water.

\begin{figure}[!htb]
\centering
\includegraphics[width=0.45\textwidth]{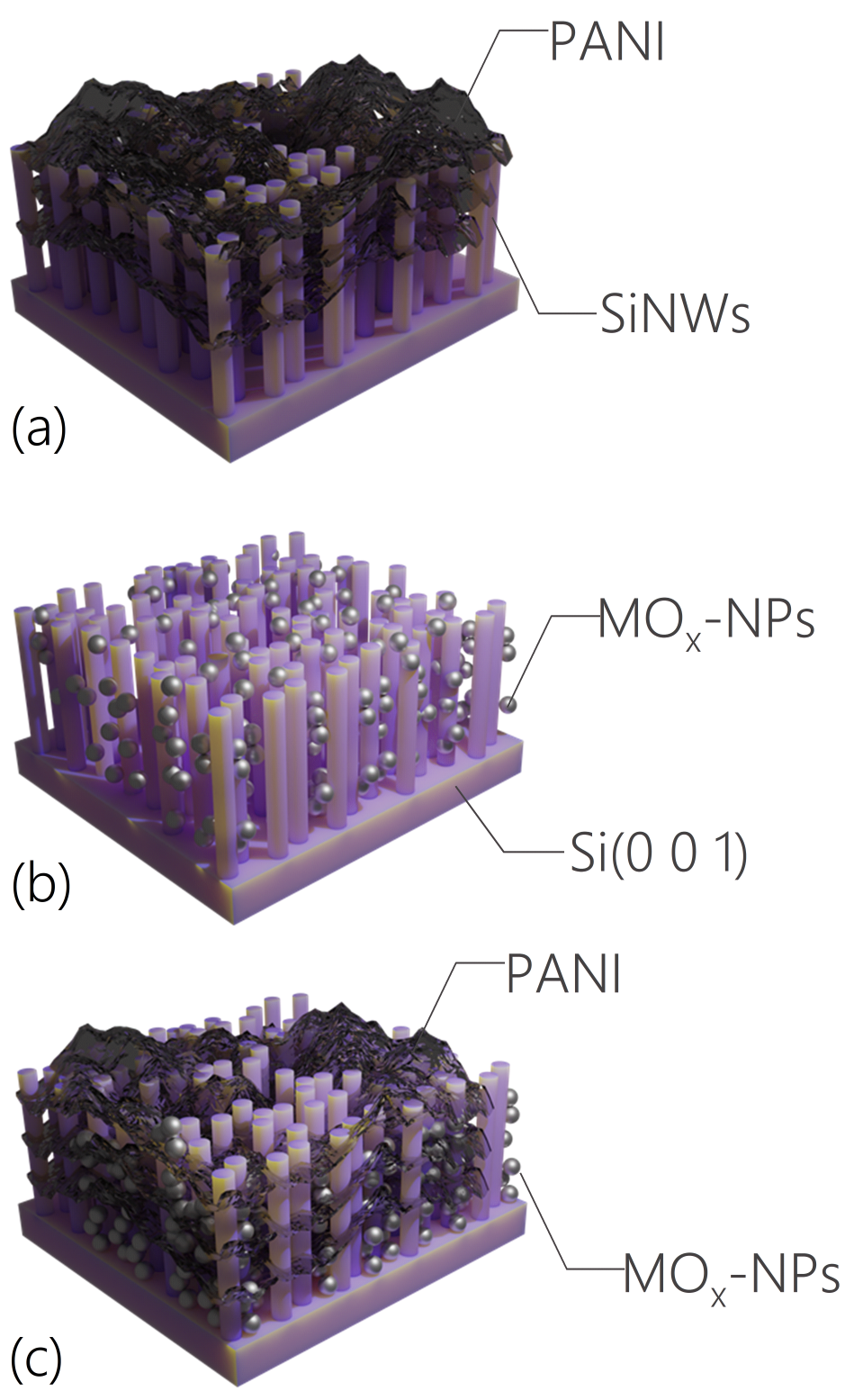}
\caption{Representation of structural schemes considered for respiratory sensing mechanism, i.e., (a) SiNWs with PANI, (b) SiNWs with MO$_x$ particles and (c) SiNWs coated with hybrid PANI:MO$_x$ nanocomposite.} 
\label{2}
\end{figure}

\item Etching the resulting sample from the previous step in HF:H$_{2}$O$_{2}$ (5M:0.4M) solution for 20 min to obtain vertically aligned SiNWs. The etching was abrupted by immersing the sample in DI-water.

\item The resulting structure is immersed for few seconds in 60\% of nitric acid to remove residual Ag-NPs, and rinsed afterwards with DI-water.

\end{itemize}

Various schemes of structures are considered in this study (see Fig. \ref{2} for schematics), one in which MO$_x$-NPs are  spin coated over the SiNWs, later consisting of PANI deposition on SiNWs and the other where the SiNWs decorated with polyaniline  MO$_x$ hybrid nanocomposite. \\

%\subsubsection{SiNWs decorated with MO$_x$-NPs}
\noindent{\bf SiNWs decorated with MO$_x$-NPs}\\

TiO$_2$ (a mixture of anatase and rutile phases, with particle size $<$ 100 nm) and WO$_3$ (particle size of $\approx 100$ nm) nanopowders from Sigma Aldrich were used for deposition of MO$_x$-NPs onto SiNWs structures. Initially the nanopowders were dispersed in DMF solvent (0.01 g/ml) and then  the solution was sonicated for an hour. A spin coater was used to coat the SiNWs with prepared nanoparticles 
 suspension. The coated  sample was placed on a hot-plate at 90$^{\circ}$C for 10 min to evaporate the solvent.\\

\begin{figure}[!htb]
\centering
\includegraphics[width=0.54\textwidth]{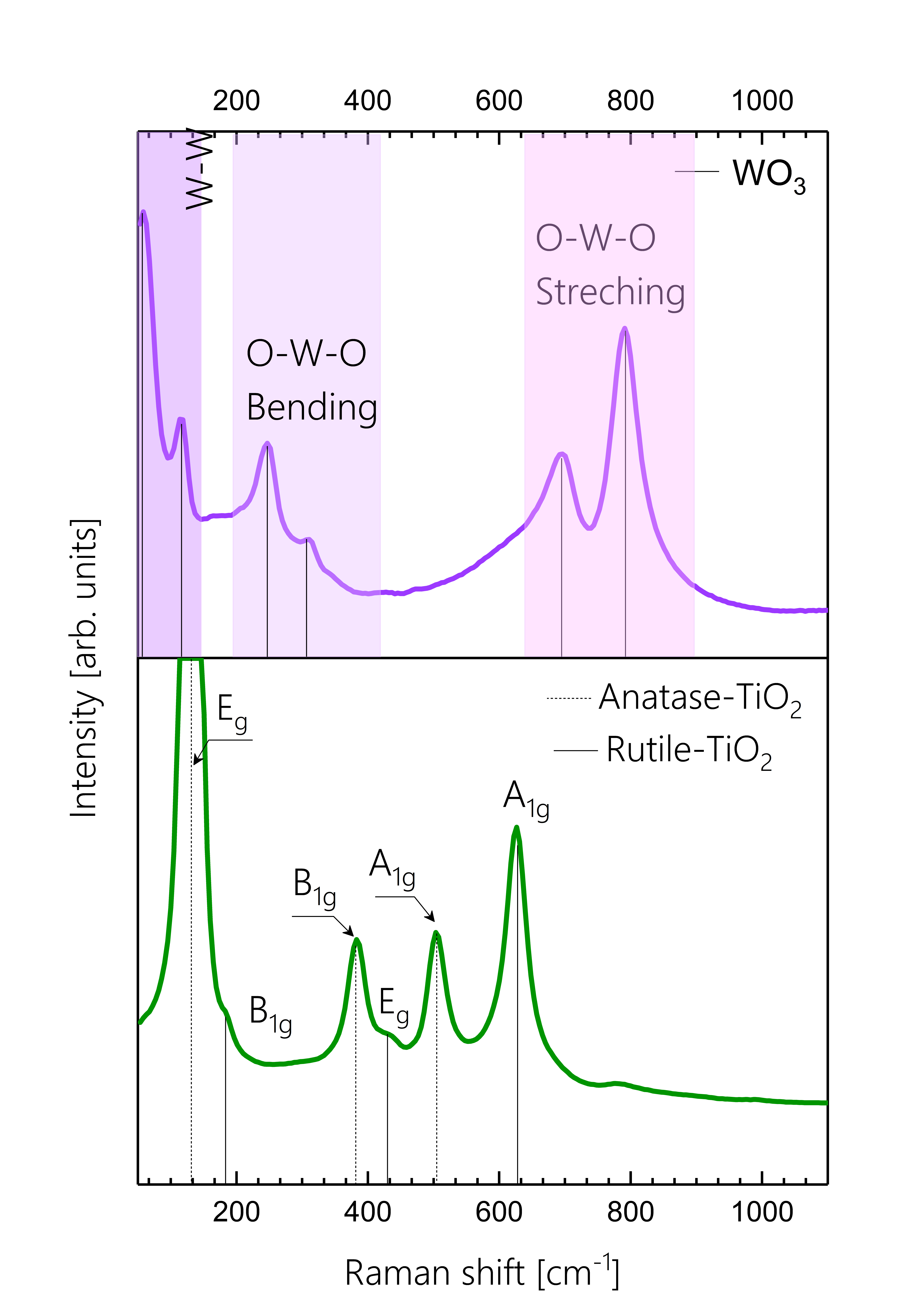}
\caption{Room temperature Raman spectra of WO$_3$ and TiO$_2$ nanopowder. The respetive peaks are marked for both WO$_3$ and TiO$_2$, respectively. In (b) the dotted line represents the Raman modes of anatase TiO$_2$ whereas the solid line represents rutile TiO$_2$.}
\label{3}
\end{figure}

\begin{figure}[!hb]
\centering
\includegraphics[width=0.48\textwidth]{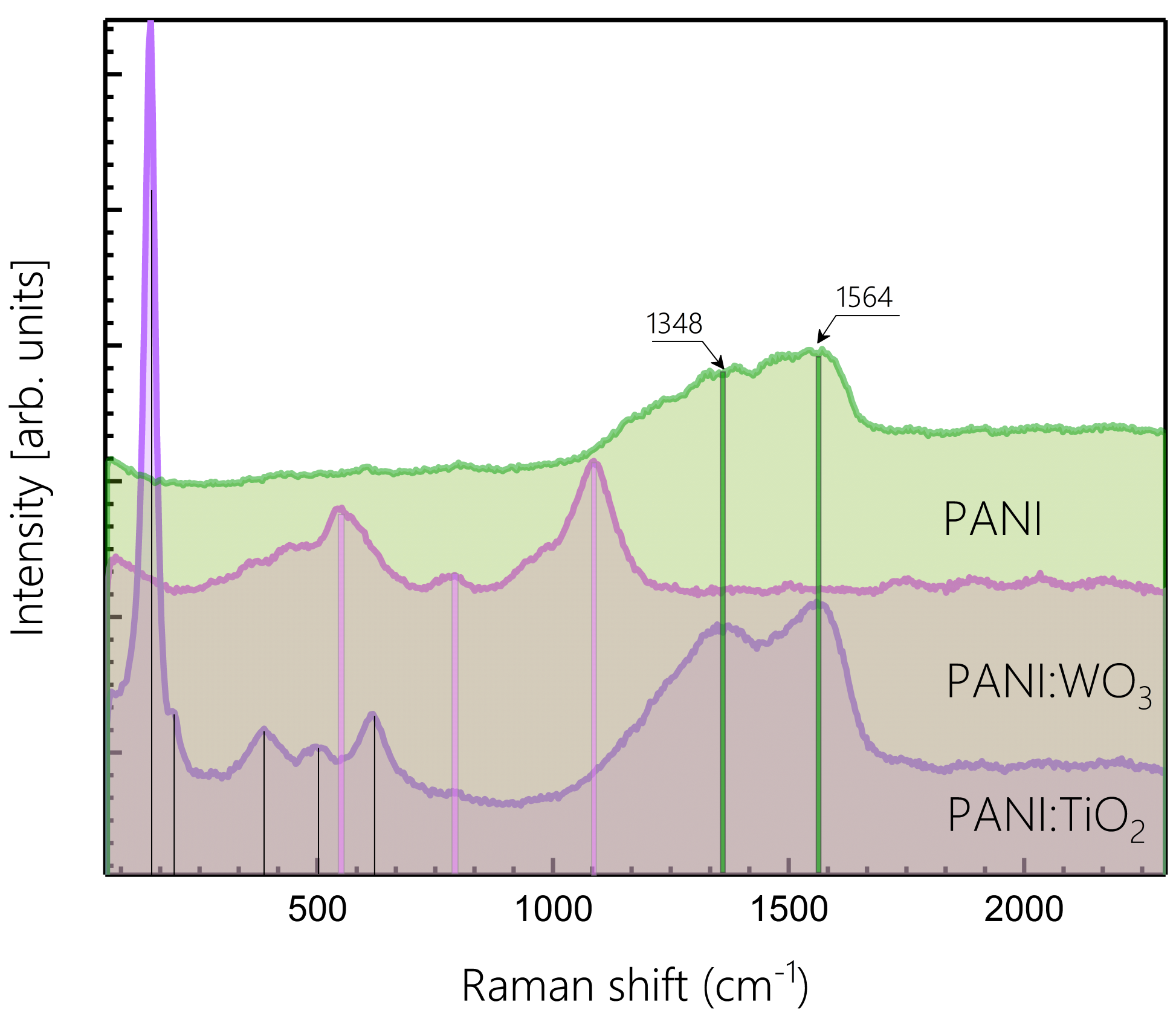}
\caption{Room temperature Raman spectra of SiNWs structure coated with PANI, PANI:WO$_3$ and PANI:TiO$_2$. The Raman peaks are marked for PANI and that for WO$_3$ and TiO$_2$, respectively.}
\label{4}
\end{figure}

%\subsubsection{SiNWs with polyaniline functionalized with MO NPS - heterostructure}
\noindent{\bf SiNWs decorated with with PANI:MO$_x$ hybrid nanocomposite  - heterostructure}\\

PANI and PANI:MO$_x$ nanocomposites were deposited on SiNWs by suspending the substrate in the polymerization solution, during  chemical oxidative polymerization of aniline with ammonium persulfate (APS) in acidic medium  \cite{Lascu2022}. For PANI deposition, SiNWs were immersed in the solution containing 0.1 M aniline in 0.1 M H$_2$SO$_4$ and kept for 30 min under continuous stirring conditions. A pre-cooled solution of 0.1M APS in 0.1M H$_2$SO$_4$ was added drop by drop to the above obtained previous solution containing the monomer. The polymerization solution was allowed to react for 24 h at room temperature under continuous stirring condition. Both, the resulting precipitate and SiNWs were collected and washed several times with deionized water and methanol. The collected powder was dried at 60 C overnight while the SiNWs decorated with PANI  was dried at room atmosphere. Similar procedure was used for deposition of PANI:MO$_x$ nanocomposites on SiNWs with the addition of  0.28 g of TiO$_2$ or 2.31 g of WO$_3$ in the monomer solution. An additional glass substrate was placed alongside the sample, for Raman spectroscopy and XRD analysis.

\subsection{Characterization}
Structural and elemental characterization of the films was conducted using X-ray diffraction Raman spectroscopy (Horiba LabRam Evolution), Scanning electron microscopy(SEM) and Energy dispersive X-ray spectroscopy (EDX) (Zeiss Supra 35). Empyrean diffractometer by Panalytical was utilised for XRD in a parallel beam geometry with a line-focused copper anode source operating at 45 kV and 40 mA with radiation Cu-K$\alpha$ (wavelength of 1.54 \AA). A parabolic x-ray mirror was used with a 1/2$\degree$ divergence slit to limit the x-ray spot size on the sample. a parallel plate collimator slit (0.27$\degree$) was used in the diffracted beam path followed by a PIXcel detector operating in open detector mode for XRD and in Frame-based mode for reciprocal space mapping(RSM). The X-ray diffraction measurement was performed for 2$\theta$ selected scanning range i.e., 10$\degree$ - 50$\degree$ for PANI coated structures and 20$\degree$ - 80$\degree$ for PANI:MO$_x$ structures. Where as, $\omega$:2$\theta$ with scanning range of 4$\degree$ was performed for RSM measurements. Raman spectroscopy analysis was made using an air-cooled frequency-oubled Nd:Yag (100 mW / 1 MHz) with 532 nm excitation line along with an 1800 mm high resolution grating. Each Raman measurement was composed of 20 integrated spectra with an acquisition time of 10 s each. The laser power setting was adjusted to 10\% in order to avoid any heating effects during measurements. The electrical characterization setup contains a source meter (Keithley 2400), a controlled sample stage with a micormanipulator, vacuum suction and stainless-steel contact arms connected to source meter and a humidity sensor.

%%%%%%%%%%%%%%%%%%%%
\section{Results and discussion}
%%%%%%%%%%%%%%%%%%%%

%%%%%%%%%%%%%%%%%%%%%%%
\subsection{Structure characterization}
%%%%%%%%%%%%%%%%%%%%%%%
Prior to deposition, MO$_x$ nanoparticles (Sigma Aldrich) were characterized by Raman spectroscopy (see Fig. \ref{3} ). The Raman spectra of  WO$_3$ nanoparticles showed well-resolved peaks at 248 and 310cm$^-1$ ascribed to O-W-O bending and at 695.8 and 789.5cm$^-1$ attributed to  O-W-O stretching  \cite{Xu2014}. As expected the Raman spectra of TiO$_2$ nanoparticles consists of a mixtures of rutile and anatase phases, as indicated by the solid and dashed lines, respectively. Three characteristics active modes positioned i.e., E$_g$, B$_{1G}$ and A$_{1G}$ for both \textit{a} and \textit{r}-TiO$_2$ observed are inline with other studies \cite{Challagulla2017, Hardcastle2011}.

\begin{figure}[!htb]
\centering
\includegraphics[width=0.5\textwidth]{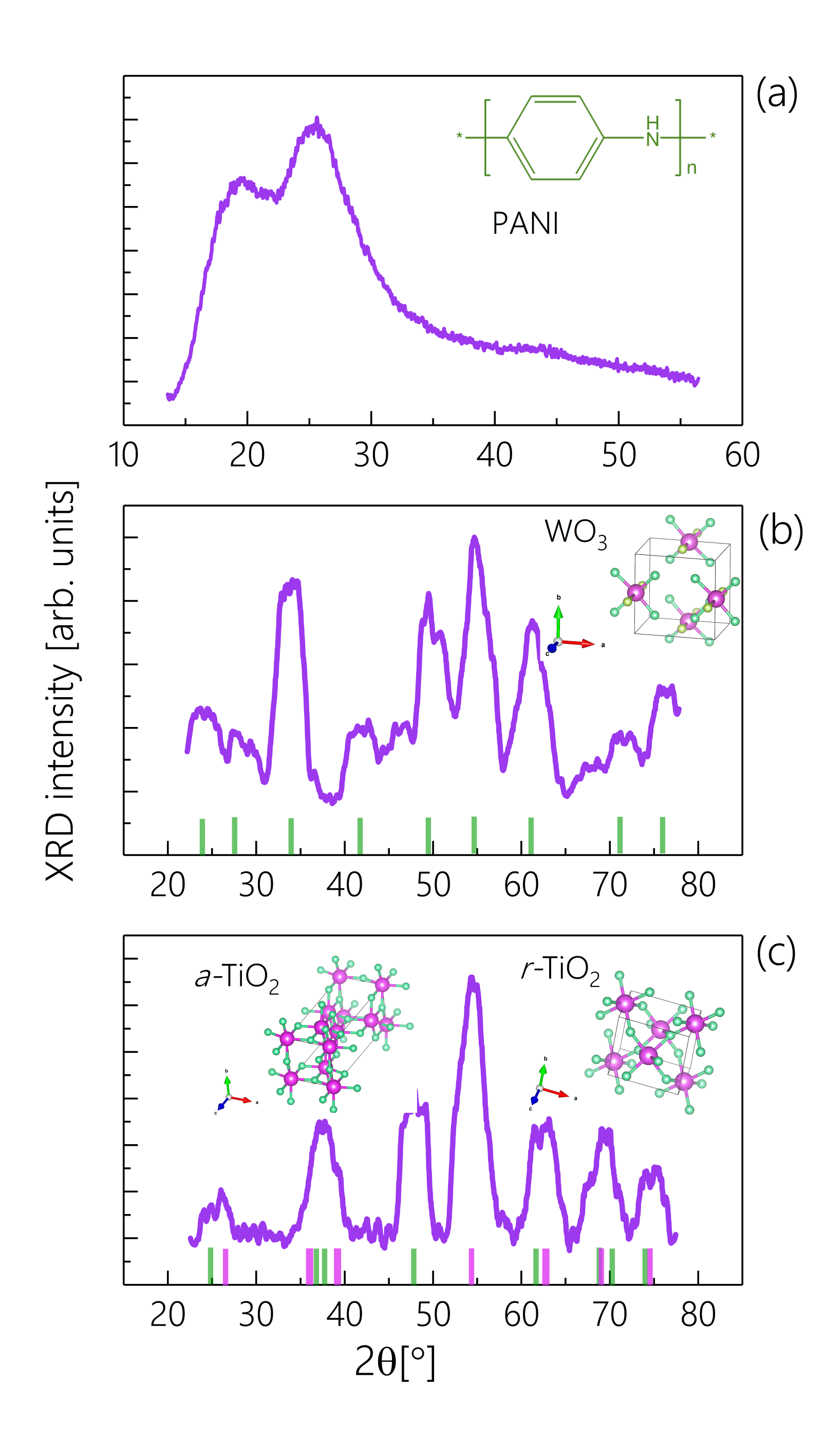}
\caption{XRD diffractogram for (a) polyaniline (b) WO$_3$ nanopowder and TiO$_2$ nanopowder. The green indicated line in (b, c) represents the standard tabulated position of WO$_3$ and anatase (\textit{a}-TiO$_2$) according to JACPD card no.: , and the red indicated line in (c) represents the standard tabulated position for rutile(\textit{r})-TiO$_2$, according to JCPDS no. 98-007-1692,00-021-1272 and 00-021-1276, respectively. }
\label{5}
\end{figure}

\begin{figure}[!htb]
\centering
\includegraphics[width=0.5\textwidth]{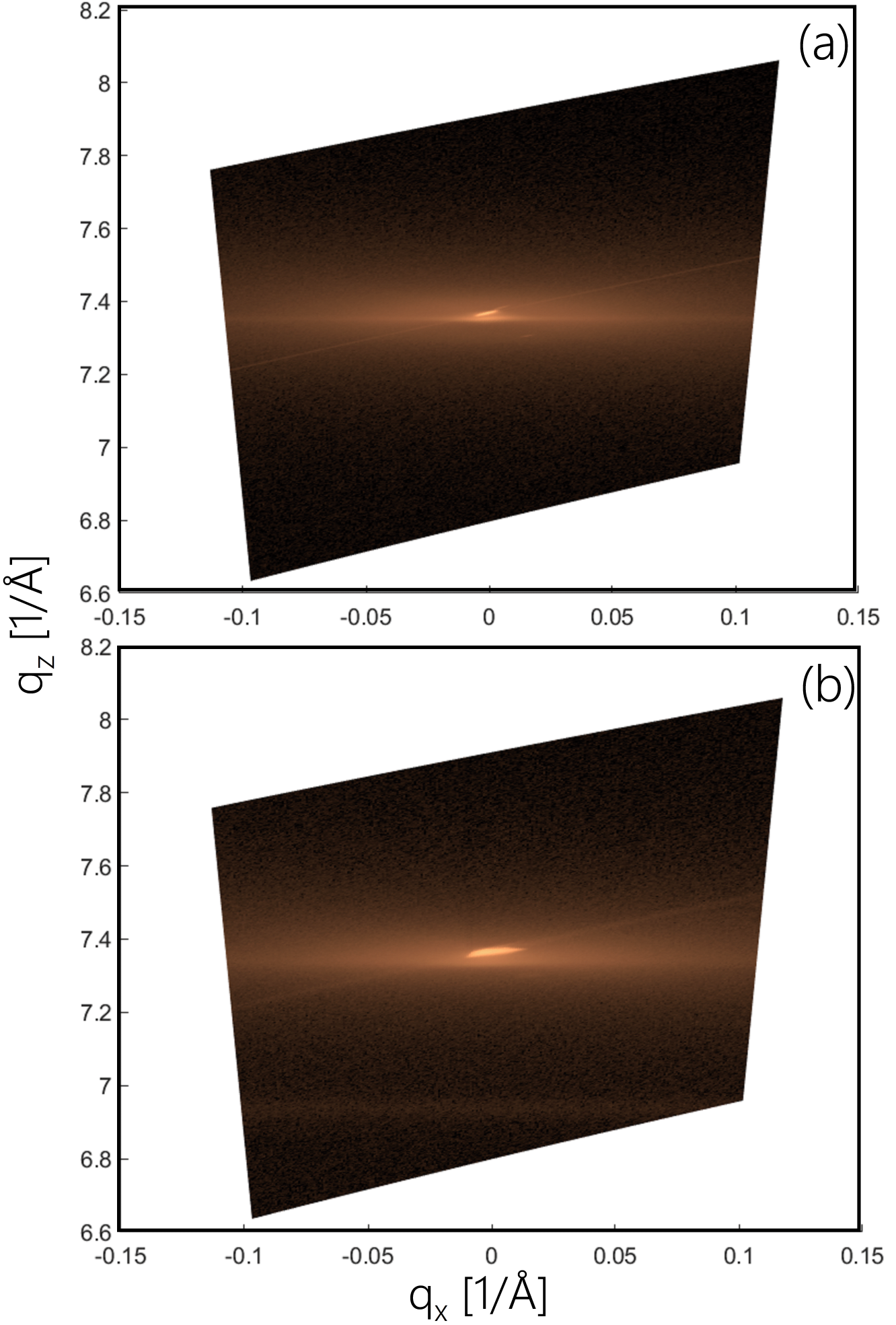}
\caption{Reciprocal space map of SiNWs and SiNws with PANI along (0 0 4) crystallographic plane.}
\label{6}
\end{figure}

Room temperature Raman spectra of PANI, PANI:WO$_3$ and PANI-TiO$_2$ structures are presented in Figure. \ref{4}.  The representative Raman spectra of structures were deconvoluted by Gaussian fitting to determine the characteristics raman shift positions. For PANI, the vibration modes (marked in figure for reference) observed at 1564 cm$^{-1}$ corresponds to  C=C stretching vibration of the quinoid ring and 1348 cm$^{-1}$ is associated with C–N stretching vibration of semiquinone radicals, being  an indication of the synthesis of the conductive form of PANI   \cite{Rajagopalan2015}. A room-temperature Raman spectra of PANI:MO$_x$ is shown in the same figure. An evident presence of characteristic peaks associated to WO$_3$ and TiO$_2$, respectively in relation with those shown in Fig. \ref{3}, was observed along with the simultaneous presence of peaks attributed to PANI.

The structure analysis for SiNWs/PANI, SiNWs/PANI:WO$_3$, SiNWs/PANI:TiO$_2$ were further conducted using grazing incidence X-ray diffraction and shown in Fig. \ref{5}. Due to the perpendicular and parallel periodicity of PANI chains, XRD pattern of PANI exhibit two broad peaks located  at 19.4$\degree$ and 25.47$\degree$ 2$\theta$ values  attributed to (020) and (200) crystallographic planes of PANI  \cite{Amaechi2015,Ambalagi2018,Lascu2022}. For WO$_3$, \textit{r} and \textit{a}-TiO$_2$ the peaks are positioned at standard tabulated values according JCPDS no. 98-007-1692, 00-021-1276 and 00-021-1272, respectively \cite{Challagulla2017, Hardcastle2011}. For TiO$_2$ the crystallographic planes located at marked 2$\theta$ values belongs to (110), (101), (211), (002), (112), (320) and (101), (103), (004), (112), (002), (213), (116), (220), (107) for rutile and anatse, respectively. Where as, for WO$_3$ the planes are attributed to monoclinic WO$_3$ i.e., (002),(022), (122),(211), (114), (213), (053) and (153) \cite{PAKDEL202330501}.

Additionally to investigate the structural features of SiNWs, X-ray reciprocal space maps (RSMs) around Si (004) reciprocal lattice point was performed giving information regarding the out-of-plane lattice, strain and the crystal imperfections \cite{Fewster1997}. X-ray RSM along (q$_z$, q$_x$) coordinates for SiNWs alone and those coated with PANI are presented in Fig. \ref{6}(a, b).

The q$_x$ and q$_z$ coordinates are projections of the scattering vector along [100] and [001] directions, and are given as: \[ q_x = \frac{2 \sin(\omega - \theta)}{\lambda} \] and \[ q_z = \frac{2 \sin \theta}{\lambda} \]

For SiNWs an intense peak located around q$_z$ $\in$ (0.7360 - 0.7365) {\AA}$^{-1}$. For cubic crystal the lattice constant a can be expressed as 4/q$_z$ \cite{Stanchu2017}, and is calculated to be 5.43 {\AA}, which corresponds to the lattice parameter of bulk Si. This sructinizes that the MACE process to obtain SiNWs does not affect the value of the lattice parameter of the samples. For SiNWs coated with PANI, the spot broadening increases in both q$_z$ and q$_x$ direction, and can be ascribed to bending and torsion acting on NWs, as a consequence of higher surface energy. Furthermore, the q$_x$ area elongation in RSM for structures which is related to diffuse scattering is observed and can be associated to crystal imperfections. Whereas a wider angular dispersion for SiNWs/PANI is observed.

\subsection{Scanning electron microscopy analysis}
\begin{figure*}%[!ht]
\centering
\includegraphics[width=0.9\textwidth]{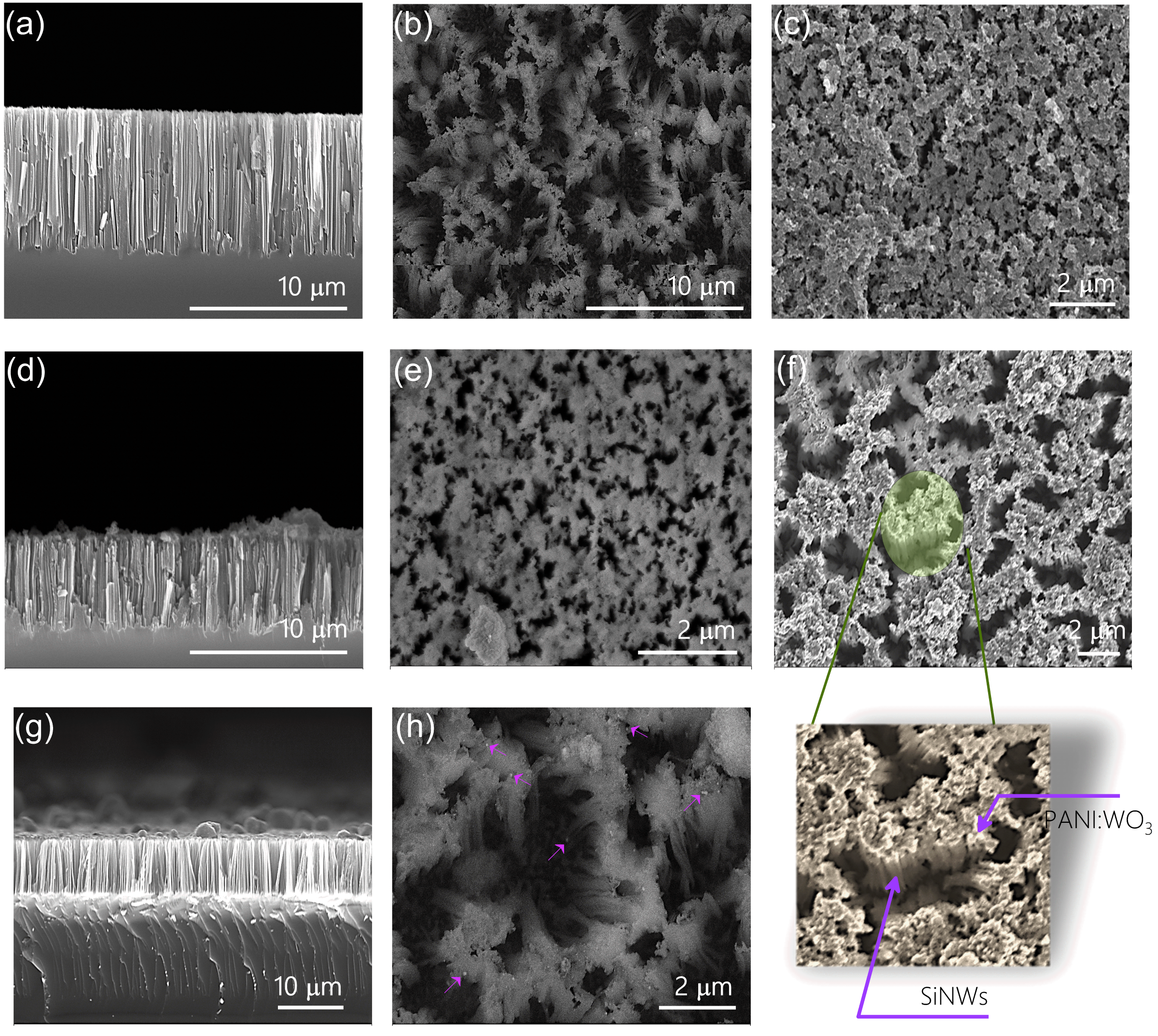}
\caption{SEM microgrpahs (a,b) cross-sectional and top-view of SiNWs structures obtained by MACE, (c) top-view of SiNWs coated with PANI, (d, e) cross-sectional and top-view of SiNWs coated with PANI:TiO$_2$, (f) top-view of SiNWs coated with PANI:WO$_3$, along with a magnified image showing the presence of NWs and (g, h)cross-sectional and top-view of SiNWs spin-coated with WO$_3$, respectively. The scale bars are provided along with each figure. }
\label{7}
\end{figure*}

Figure \ref{7} shows the top and cross-section-view of the SiNWs structures both without and those coated with PANI, PANI:MO$_x$ and  spin coated with MO$_x$-NPs. The length of nanowires measured was approximately 6.5$\mu$m with diameter ranging between 20 to 50 nm (Fig. \ref{7}(a)) depending of size distribution of Ag particles obatined during deposition using AgNO$_3$. For SiNWs (Fig. \ref{7}(a, b), the SEM image revealed vertically aligned randomly distributed wires, bundled together due to capillary effect induced by wetting as is evident in Fig. \ref{7}(b). For structures coated with PANI (Fig. \ref{7}(c)), the top-view showed a cauliflower morphology of PANI, and is similar to those obtained in study by Lascu \textit{et al}.  \cite{Lascu2022}.

Figure. \ref{7}(d, e, f) shows cross-sectional and top view of PANI:TiO$_2$ and PANI:WO$_3$ structures. It was observed that compared to PANI:WO$_3$, SiNWs coated with PANI:TiO$_2$ showed agglomertaion of TiO$_2$ nanoparticles over the surface of PANI. Further, deposition of PANI:MO$_x$ on SiNWs resulted in increasing surface coverage, i.e., possible increase in NWs bundling as is supported by RSM plot showing increased q$_x$ value.

For SiNWs structures spin-coated with TiO$_2$ nanoparticles, SEM image are shown in Fig. \ref{7}(g, h). An agglomerated TiO$_2$ nanoparticles can be seen in the surface whereas from top view one can visulize smaller particles went with in the nanowires. 

The EDS analysis of the structure are shown supplementary information (Fig. S1). The obtained results scrutizinizes the presence of PANI and nanoparticle with highest intensity attributed to be originated from the Si-substrate.

\subsection{Electrical Characterization}

Before we delve in to the electrical characterization of SiNwS its important to understand the sensing mechanism of interconnected silicon nanowires (SiNWs) which operates as follows: to detect respiration, there must be an interaction between moisture and SiNWs, leading to moisture absorption onto the SiNWs' surface, facilitated by the highly hydrophilic nature of the nanowires \cite{Song2023, Akbari}. This phenomenon induces changes in the SiNWs' hole accumulation layer(HAL) width and surface potential, consequently altering the conductance of the SiNWs. 
Since moisture itself is not inherently oxidizing or reducing in this context, in the presence of an oxidizing gas, the gas extracts electrons from the SiNWs' conduction band, widening the HAL width \cite{Qin2018, Morganti2021}. Conversely, in the presence of a reducing agent, electrons are released and trapped by oxygen molecules, resulting in a narrowing of the HAL width \cite{Qin2018}.

\begin{figure*}[htb]
\centering
\includegraphics[width=1\textwidth]{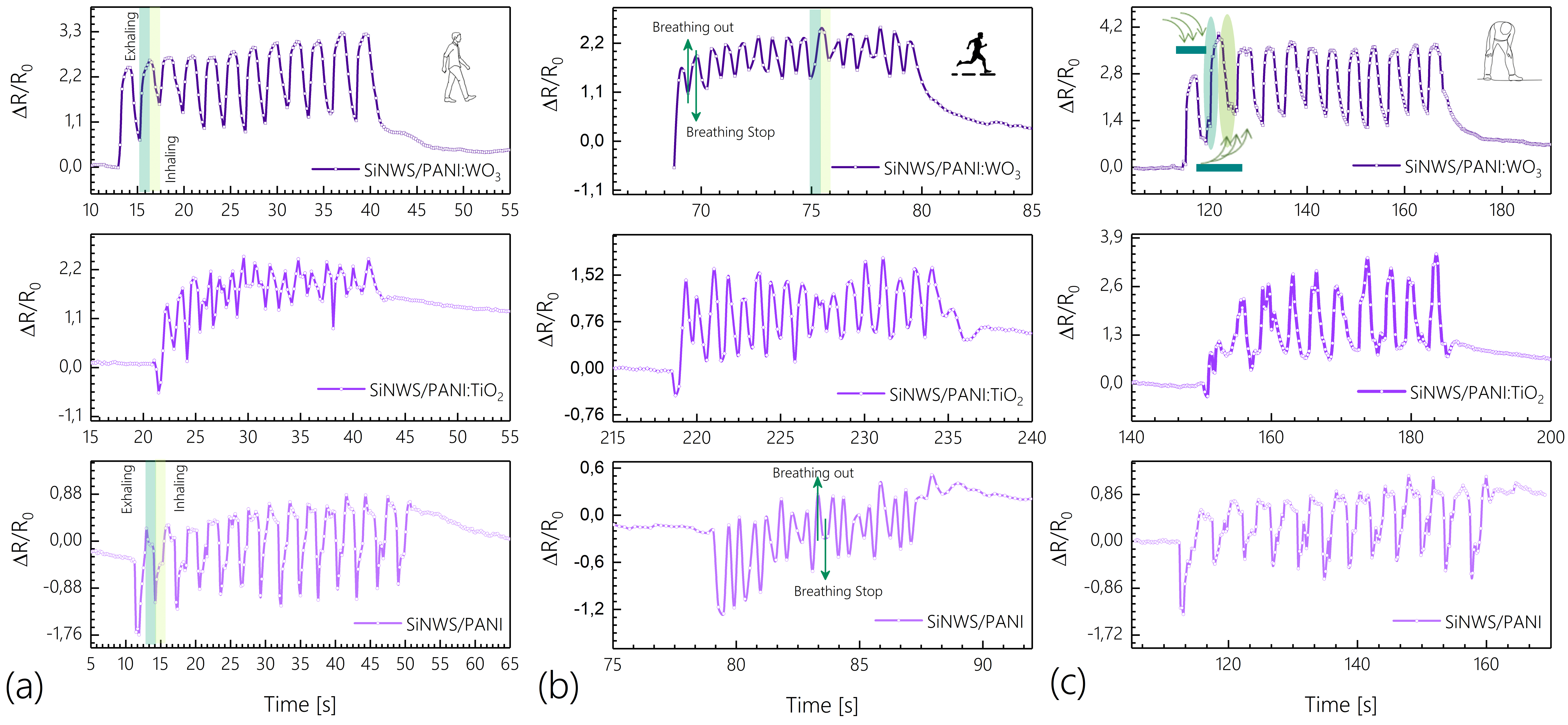}
\caption{Respiratory sensing for three different breathing patterns i.e., (a-c) normal, rapid and deep breathing, respectively, using SiNWs structures decorated with PANI and PANI:MO$_x$, respectively. The highlighted regions represents the exhaling and inhaling stimulus characteristics.}
\label{8}
\end{figure*}

Expanding on this, moisture in the breath ionizes into hydronium and hydroxide ions, establishing an electric double layer between the moisture and the SiNWs \cite{Song2023,Akbari}. As moisture permeates the nanochannels, it induces electron accumulation in the silicon, causing them to migrate towards the anode in association with holes generated at the cathode. The unique forest-like morphology of SiNWs, coupled with their large surface area, significantly influences the sensor's response \cite{Akbari}. In our previous works we have studied the respiratory sensing using SiNWs obtained by MACE which have shown reasonable response to human breath. In this study we intended to enhance the respiratory sensing properties using SiNWs decorated with PANI and  PANI:metal oxide hybrid nanocomposites.

The electrical characterization of structures was carried out using custom-built setup. For respiratory sensing, the distance between the sample and the human nose was $\sim$3mm. Three often used respiratory profiles i.e., normal breathing (NB), Tachypnea (FB-usually more than 20 breaths/minute  \cite{Park2024}) and deep breathing (DB) were monitored. The data acquisition speed was set to $\sim$1 ms (limited by the equipment). 

Figure \ref{8}. shows the sensitivity of structures obtained by change in resistance under various respiratory patterns obtained over SiNWs structures coated with PANI, PANI:WO$_3$ and PANI:TiO$_2$. The sensitivity and response time of the structures increases with the introduction of MO$_x$ nanoparticles in PANI i.e., PANI$<$PANI:TiO$_2$$<$PANI:WO$_3$. The response time $t$ is given as:

\begin{equation}
    R_{\textit{t}} = t_{90\%} - t_{10\%},
\end{equation}

where, t$_{90\%}$ and  t$_{10\%}$ are the time when the change due to external stimuli is at 90\% and at 10\% respectively. The R$_t$ calculated are tabulated in table \ref{table:1} and is an average value calculated over several waveform. Furthermore, compared to SiNWs structure coated with PANI:WO$_3$, the structures coated with PANI and PANI:TiO$_2$ showed distorted features, along with a significant drift in baseline, regarding SiNWs coated with PANI:WO$_3$ to posses enhance sensitivity for respiratory sensing. It is important to point out that the profiles observed for SiNWs coated with PANI showed rather an opposite change i.e., drop in resistance during exhaling, contrary to an increase in resistance for structures coated with PANI:MO$_x$. Such a behaviour can be understand by understanding the sensing mechanism in respective structure discussed later.  For better visualization the exhaling and inhaling are highlighted in Fig. \ref{8}.

\begin{table}[h!]
\centering
\begin{tabular}{|c|ccc|}
\hline
\textbf{Structures} & \multicolumn{3}{c|}{\textbf{R$_t$[s]}} \\ \hline
\textbf{} & \textbf{NB} & \textbf{FB} & \textbf{DB} \\ \hline
SiNWs/PANI & 1.32 & 0.3 & 1.63 \\ \hline
SiNWs/PANI:TiO$_2$ & 0.44 & 0.21 & 0.35 \\ \hline
SiNWs/PANI:WO$_3$ & 0.38 & 0.15 & 0.33 \\ \hline
\end{tabular}
\caption{Response time calculated from Fig. \ref{8} for respective structures.}
\label{table:1}
\end{table}

The sensing mechanism in Polyaniline (PANI) involves protonation and deprotonation processes facilitated by the adsorption and desorption of moisture \cite{Kundu2020}. In PANI, various protonic sites are available, which act as proton donors to water molecules, leading to the formation of oxidized sites (-NH$^+$=) \cite{Khuspe2012, Morsy2022, Cavallo2015}. This interaction involves water molecules accepting protons to form hydronium ions \cite{Morsy2022, Kundu2020}, which subsequently undergo proton hopping between adjacent water molecules \cite{Fratoddi2015, Morsy2022, Kundu2020}, following the Grotthuss mechanism. In terms of electron transport within PANI, as water molecules accept protons, electron rearrangements occur. For instance, during inhalation, lone-pair electrons from nitrogen atoms donate to protons, forming -NH$_2$$^+$- groups within PANI. Conversely, during exhalation, water molecules withdraw protons, resulting in the release of lone-pair electrons from nitrogen atoms, which can then participate in the resonance of phenyl rings in PANI. This resonance delocalization of electrons leads to either an increase in conductivity or a decrease in resistance within PANI.

\begin{figure}[!htb]
\centering
\includegraphics[width=0.42\textwidth]{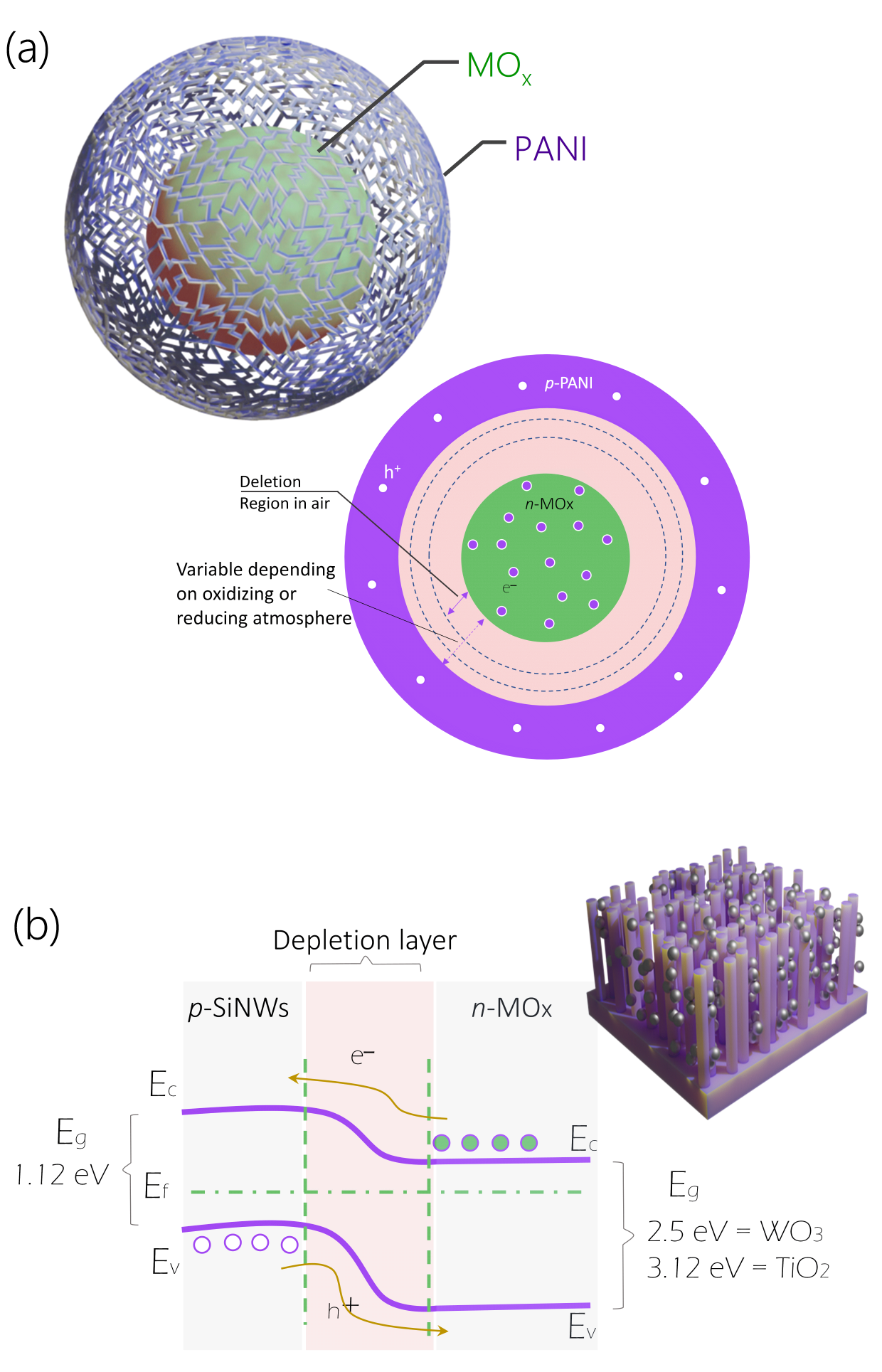}
\caption{Schematic illustration of (a) p-n junction formation in (a) SiNWs coated with hybrid p-type PANI encapsulting n-type MO$_x$ nanoparticles. The Figure demonstrate the depletion layer width alteration in air and in presence of oxidzing or reducing agent. and (b) energy band with p-n junction formed between p-type SiNWs spin coated with n-type MO$_x$ nanoparticles.}
\label{10}
\end{figure}

Contrary to PANI, a rather interesting mechanism takes place in hybrid PANI/MO$_x$ coated SiNWs structures. There the sensing is predominantly based on the trapping of moisture in hybrid nano-composite structure resulting in  change in the surface resistance of the nanocomposites when the molecular species adsorb and react with it.  Whereas, in pure metal-oxides coated SiNWs the PANI/MO$_x$ coated structures are more porous as a consequence of MO$_x$-nanoparticles in PANI matrix, providing active centrers giving access to localized donor and acceptor states \cite{Kulkarni2019}. An interesting aspect of this structure is the formation of p-n hetero-junction which has formed between p-type PANI and n-type MO$_x$ particles \cite{Kulkarni2019, Tian2016}.  When the nanocomposite was exposed to moisture, due to electron and hole transfer between PANI and WO$_3$, a depletion layer is formed at the interface of PANI and WO$_3$, resulting into formation of hetero-junction barrier \cite{He2020, Kulkarni2019} (shown in Fig. \ref{10}). PANI is a p-type material, since majority charge carriers are holes. When the hybrid structure is exposed to moisture at room temperature, the free electrons generated neutralize the holes due to electron-hole combination resulting in decrease in the hole concentration in PANI. This reduces the carrier concentration of the hetero-junction and potential barrier height increases followed by increase in the resistance of hybrid structure \cite{Tian2016, Kulkarni2019}.

\begin{figure}[htb]
\centering
\includegraphics[width=0.48\textwidth]{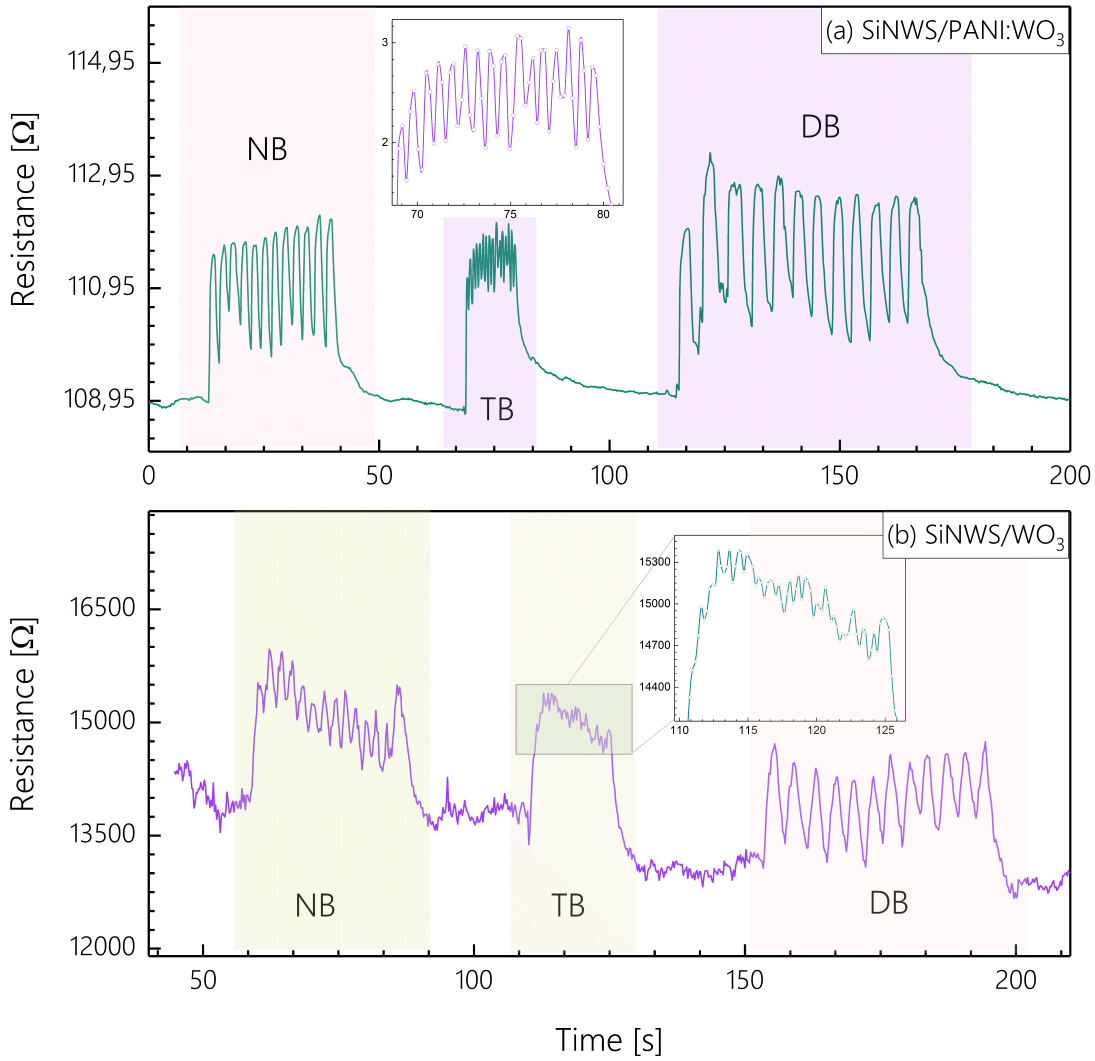}
\caption{Plot for room temperature change in resistance as a function of time under varying breathing stimulus using SiNWs structures decorated with (a) PANI:WO$_{3}$ using electroless deposition and (b) WO$_3$ nanoparticles using spin-coating, respectively. The insets show magnified pattern under rapid breathing. }
\label{9}
\end{figure}

Additionally a comparison between SiNWs structures decorated with PANI:WO$_3$ and those spin-coated with WO$_3$ nanoparticles was made (Fig. \ref{9}). It was observed that the introduction of semiconducting polymer composite with WO$_3$ exhibited lower resistance and  shorter response time(R$_{\textit{t}}$). The R$_t$ calculated for SiNWs structure decorated with PANI/WO$_3$ and WO$_3$-NPs is on average 0.607, 0.1, 0.96 s and 0.99, 0.33, 1.62 s for NB, FB and DB respiratory pattern respectively. Such relative enhancement in response time can be attributed to uniform distribution and adherence of PANI:WO$_3$ between interconnected nanowires, whereas, as evident by SEM the spin coated WO$_3$ results in agglomeration of particles over the surface while nanoparticles those present within the nanowires are spread far from each other with less adherence and fictionalization properties.

\begin{figure}[htb]
\centering
\includegraphics[width=0.45\textwidth]{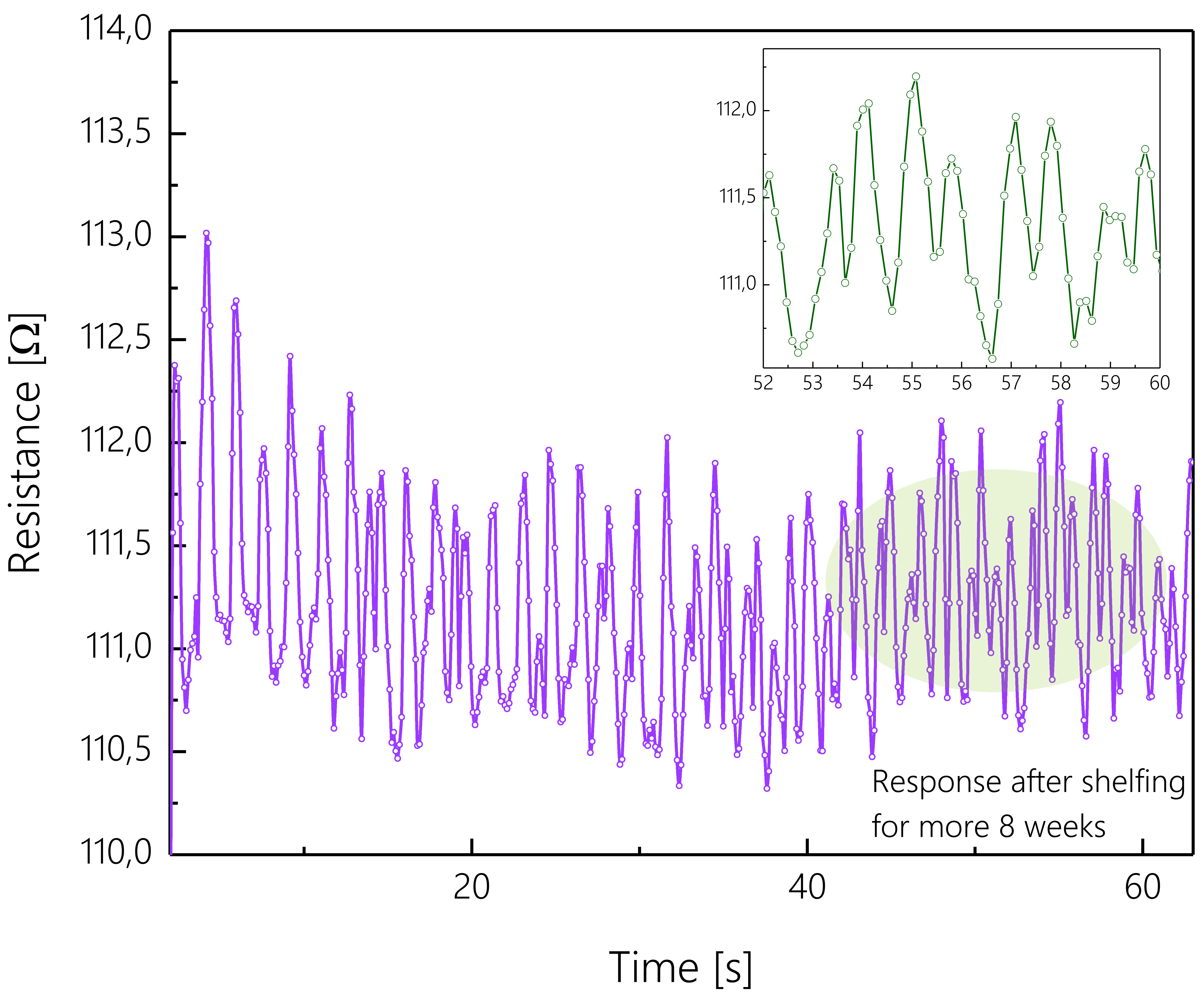}
\caption{Response of the structure (SiNWs/ PANI: WO$_3$) after shelving for more than two-months. }
\label{11}
\end{figure}

\begin{figure*}[!htb]
\centering
\includegraphics[width=1\textwidth]{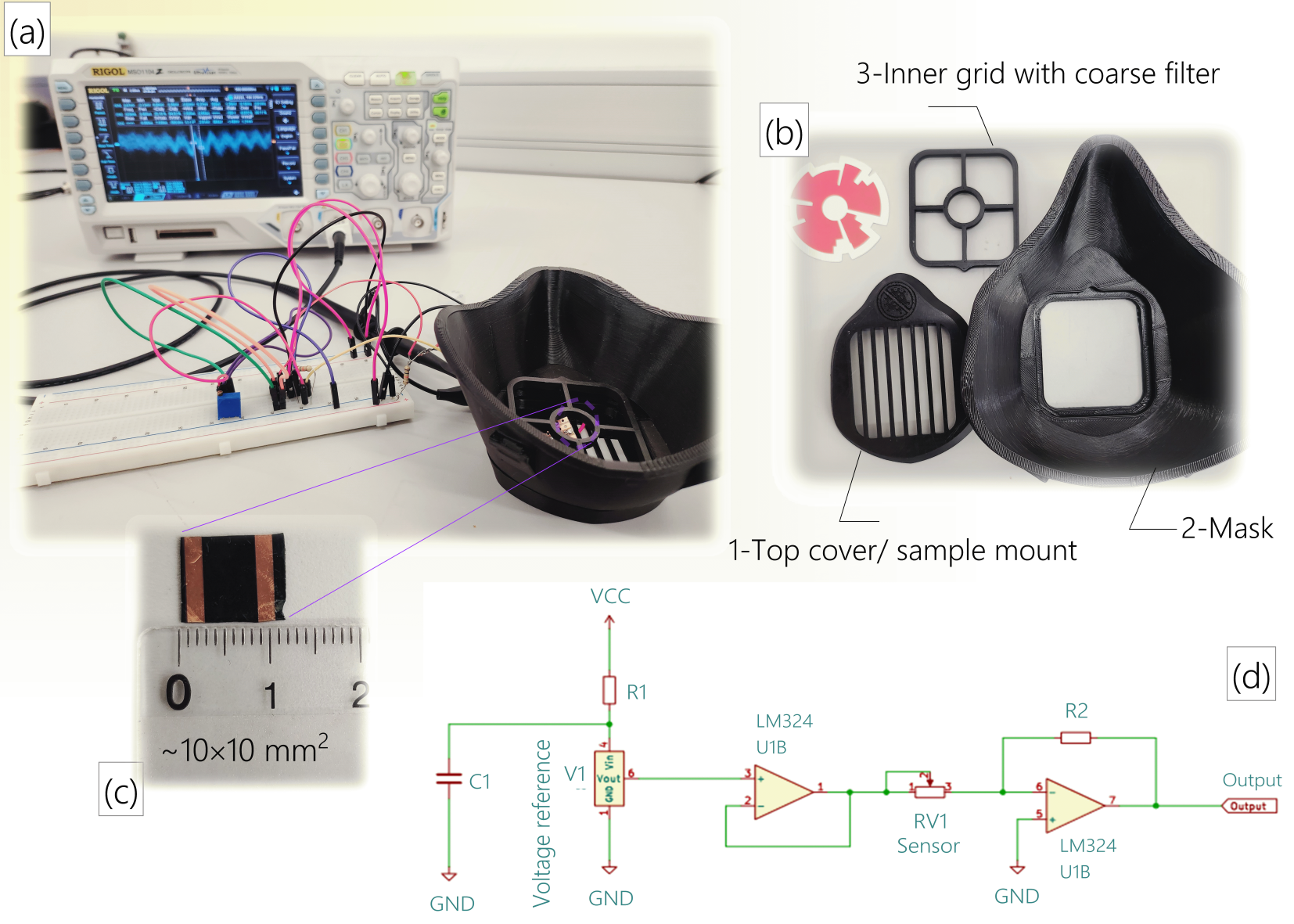}
\caption{(a) Photograph of data acquisition electronics, utilizing an oscilloscope, assisted with Arduino microcontroller board(not shown here). (b) 3D printed mask with a outer cover for mounting sample and an inner grid which can can integrated with course filter for damage protection. (c) Sample before mounting with size $\sim$10$\times$10 mm$^2$. (d)Circuit diagram of breath sensor. }
\label{12}
\end{figure*}

To elaborate on the sensing mechanism let us first introduce the mechanism taking place in n-type WO$_3$ and TiO$_2$ i.e., metal oxides. The sensing mechanism  is related to the adsorption and desorption of molecules on the sensor surface, causing a change in resistance. That is oxygen molecule in air reacts with electron when is adsorbed on the surface of metal oxides, resulting in decrease in electronic concentration on the surface of WO$_3$ and an increase of width of depletion regions \cite{Masuda2023, He2020} and thus results in an increase in resistance of the sensor, as is observed in our case (see Fig. \ref{10}). 

Whereas, in case of SiNWS/MO$_x$ structure mechanism can be described with initial formation of junction between the p-type SiNWs with n-type MO$_x$, playing an important role in modulation the sensing behaviours of NWs. As documented in  \cite{Akbari} dominant conductive path is the  interconnected-SiNWs, and the diffusion of charge carriers takes place at the interface of the NWs sand MO$_x$ nanoparticles due to difference in the Fermi-level \cite{Liu2017}. This results in the formation of the depletion region with an inner electric field at the interface. That is as the O$_2$ molecules are adsorbed at the surface, the electrons are extracted thereby resulting in an increase in hole concentration in p-type SiNWs thus distorting the balance in depletion layer. That is a positive charge layer hinders the diffusion of hole from p-type SiNWs to TiO$_2$, causing a reduction in depletion layer thickness \cite{Qin2017, Liu2017}.

Returning to comparison between the hybrid and MO$_x$ SiNWs structure as observed by results in Fig. \ref{9}, the hybrid structure obtained by growing cauliflower like PANI WITH WO$_3$ nanoparticles showed increased response with detailed feature and can be ascribed to large surface area than that of pure WO$_3$, favoring more molecules to assimilate on to the surface giving improved sensitivity. Additionally since acidified PANI which one of the conductive polymers has a wider conduction channel leads to a lower resistance in the structure compared to that of SiNWs/MO$_x$ as is evident by the plots in Fig. \ref{9} and similar to what was observed in study by He \textit{et. al.}  \cite{He2020}. The hybrid structure have more oxygen vacancies than pure WO$_3$, which promotes the adsorption of gas molecules.

To investigate the stability of sensor, the structure was shelved for more than 2-months, after which when measured resulted in insignificant or no change in resistance or deterioration of observed respiratory profile (Fig. \ref{11}). 

In efforts towards practical solutions for real-time monitoring of breathing patterns we devised a breath sensor (Fig. \ref{12}). The device features a mask outfitted with the sensor, shielded with a grid for optimal performance and protection to possible damages. A photograph of the circuit and circuit diagram can be visualized in the figure. The data acquisition is performed using an Oscillscope assisted with Arduino microcontroller board(not shown here) for recording the data.

%%%%%%%%%%%%%%%%%%%%
\section{Conclusion}
%%%%%%%%%%%%%%%%%%%%

In summary, we successfully synthesized a hybrid structure of PANI:MO$_x$ decorated on SiNWs, using a cost-effective and simple chemical oxidative polymerization method for their application in respiratory sensing. The structure were characterized using RAMAN, XRD and RSM and EDs analysis, confirming the presence of PANI encapsulating metal oxide nanoparticles i.e., WO$_3$ and TiO$_2$. The morphology studied using SEM micrographs showed the formation of vertically aligned randomly distributed silicon nanowires, covered with porous structure of PANI:MO$_x$. 

Comparative analysis with SiNWs coated with PANI or WO$_3$ alone showed that the PANI:MO$_x$ hybrid composite exhibited superior response in respiratory sensing for various breathing profiles at room temperature. Particularly, PANI:WO$_3$-coated structures showed enhanced sensitivity and response time with minimal or no baseline drift.

Furthermore, to scrutinize the stability and application in real-time respiratory monitoring the sensor was shelved for a period of over two. After which when tested showed no change in sensing characteristics. These findings highlight the promising prospects of hybrid PANI-based sensors in both research and industrial applications. Notably, their low fabrication cost, scalability, and ease of fabrication pave the way for the development of efficient respiratory sensing technologies.

%https://link.springer.com/article/10.1007/s00339-016-0468-y
%file:///C:/Users/muhammad16/Downloads/Thickness_and_angular_dependent_magnetic_anisotrop.pdf
%://www.researchgate.net/publication/320687271_Thickness_and_angular_dependent_magnetic_anisotropy_of_La_067_Sr_033_MnO_3_thin_films_by_Vectorial_Magneto_Optical_Kerr_Magnetometry

%%%%%%%%%%%%%%%%%%%%
\section{Acknowledgements}
%%%%%%%%%%%%%%%%%%%%
This work was supported by the Icelandic research fund, grant no. 239987-051, and Landsvirkjun-The Energy Research Fund grant no.: NÝR-29-2024. We greatly appreciate and would like to acknowledge the efforts of Hannes Pall Thordarson Thorfor his assistance and guidance in developing a circuit and 3D model for the respiratory sensing device.\\\par

%%%%%%%%%%%%%%%%%%%%
%\section{References}
%%%%%%%%%%%%%%%%%%%%

%\bibliographystyle{apsrev}
%\bibliographystyle{unsrt}
%\bibliography{export}

\end{document}